\documentclass[11pt]{article}
\usepackage{latexsym}
\usepackage{graphicx}
\usepackage{amsfonts,amssymb}

\begin{document}

\title{Hamiltonian Formalism of Game Theory}
\author{Jinshan Wu
\\ Department of Physics $\&$ Astronomy,\\
University of British Columbia, Vancouver, B.C. Canada, V6T 1Z1 }

\maketitle

\begin{abstract}
A new representation of Game Theory is developed in this paper.
State of players is represented by a density matrix, and payoff
function is a set of hermitian operators, which when applied onto
the density matrix gives the payoff of players. By this formalism,
a new way to find the equilibria of games is given by generalizing
the thermodynamical evolutionary process leading to equilibrium in
Statistical Mechanics. And in this formalism, when quantum objects
instead of classical objects are used as the objects in the game,
it naturally leads to the Traditional Quantum Game, but with a
slight difference in the definition of strategy state of players:
the probability distribution is replaced with a density matrix.
Further more, both games of correlated and independent players can
be represented in this single framework, while traditionally, they
are treated separately by Non-cooperative Game Theory and
Coalitional Game Theory. Because the density matrix is used as
state of players, besides classical correlated strategy, quantum
entangled states can also be used as strategies, which is an
entanglement of strategies between players, and it is different
with the entanglement of objects' states as in the Traditional
Quantum Game. At last, in the form of density matrix, a class of
quantum games, where the payoff matrixes are commutative, can be
reduced into classical games. In this sense, it will put the
classical game as a special case of our quantum game.
\end{abstract}


Pacs: 02.50.Le, 03.67.-a, 03.65.Yz

\newpage
\tableofcontents
\newpage

\section{Introduction}
For static non-cooperative games, theoretically Nash Theorem gives
a closed conclusion on the existence of equilibrium. However, the
way to calculate the equilibrium of general games are still open,
and so far the existing methods seem not very general.

On the other hand, in Physics, we do have similar situations and
solutions, and they provide prototypes to deal with the problem
above. In Statistical Mechanics, principally or at least
practically, we have no dynamical evolutionary process, or say, by
first principle, leading to thermal equilibrium, but the
definition of thermal equilibrium is definitely clear, and we do
have some pseudo-dynamical processes such as the Liouville-von
Neumann equation for relevant partial system\cite{kubo} or
artificial man-made evolutionary processes, such as Metropolis
Method\cite{newman}, giving the correct thermal equilibrium in a
quite general sense. Now we want to apply this experience onto
Game Theory.

The basis of such application is the same mathematical form should
be used in both Physics and Game Theory. Here, we choose density
matrix as this basis. The equilibrium of a statistical system is
defined as
\begin{equation}
P\left(\vec{x};\beta\right) = \frac{1}{Z}e^{-\beta
E\left(\vec{x}\right)},
\end{equation}
where $\vec{x}$ is the dynamical variable, usually including
general position and general momentum, and $Z$ is the normalized
constant,
\begin{equation}
Z=\int d\vec{x} e^{-\beta E\left(\vec{x}\right)}.
\end{equation}
Or for a quantum statistical system,
\begin{equation}
P\left(E_{n};\beta\right) = \frac{1}{Z}e^{-\beta E_{n}},
\end{equation}
where
\begin{equation}
Z=\sum_{n} e^{-\beta E_{n}},
\end{equation}
and $n$ is the complete good quantum number for this system. The
above definitions can be unified as an operator form, that
\begin{equation}
\hat{\rho}\left(\beta\right) = \frac{1}{Z}e^{-\beta \hat{H}},
\label{boltzman}
\end{equation}
and
\begin{equation}
Z=\int d\vec{x}d\vec{x}^{\prime} \left<\vec{x}\right|e^{-\beta
\hat{H}}\left|\vec{x}^{\prime}\right>.
\end{equation}
The way to use an operator form for classical systems may seem
wired even for readers in Physics. However, it's quite trivial if
we regard classical systems as systems always having trivial
eigenstates with eigenvalues $\vec{x}$, that
\begin{equation}
\hat{H}\left|\vec{x}\right> =
E\left(\vec{x}\right)\left|\vec{x}\right>,
\end{equation} and with normalization
\begin{equation}
\left<\vec{x}^{\prime}\left|\vec{x}\right.\right> =
\delta\left(\vec{x}-\vec{x}^{\prime}\right).
\end{equation}
Such unification of notation brings nothing new into Physics, but
in this paper, the operator-form formalism unifies all the
formulas so that easy to apply onto both classical and quantum
game. It's helpful to make an immediate application of this
notation, that the operator form of classical thermal equilibrium
can be written as
\begin{equation}
\begin{array}{ccc}
\hat{\rho}^{classical}\left(\beta\right) & = & \frac{1}{Z}\int
d\vec{x}d\vec{x}^{\prime}\left|\vec{x}^{\prime}\right>\left<\vec{x}^{\prime}\right|e^{-\beta
\hat{H}}\left|\vec{x}\right>\left<\vec{x}\right| \\ \\ & = &
\frac{1}{Z}\int
d\vec{x}d\vec{x}^{\prime}\left|\vec{x}^{\prime}\right>e^{-\beta
E\left(\vec{x}\right)}\delta\left(\vec{x}^{\prime}-\vec{x}\right)\left<\vec{x}\right|\\
\\ & = & \frac{1}{Z}\int d\vec{x}\left|\vec{x}\right>e^{-\beta
E\left(\vec{x}\right)}\left<\vec{x}\right|.
\end{array}
\label{pclassical}
\end{equation}
It's important to notice the general character that for a
classical system that only the diagonal terms exist, while for a
quantum system, it depends on the basis. Especially, for basis
$\left\{\left|x\right>\right\}$, the eigenvectors of position
vector $\hat{X}$, usually, the density matrix has the off-diagonal
terms.

Then all thermal dynamical quantities should be averaged over the
whole space from dynamical variables according to this Boltzman
distribution. However, for a many-body system, such summation over
the whole space is not an easy mission. So average over a sampling
of this distribution is used to replace the whole space summation.
The idea is if we have the dynamical process driving the system
into thermal equilibrium, then we can just start from an arbitrary
initial state, and then let the system evolute according to the
dynamical process. After some long time, we record the trajectory,
and use the states on this trajectory as samples of equilibrium
states. Then, the average can be done on those samples.
Unfortunately, principally we don't even have such dynamical
process. However, the artificial man-made evolutionary process,
such as Metropolis Method, survives this idea. The idea of
Metropolis Method\cite{newman} is to define an evolutionary
process also leading to the right thermal equilibrium not from
first principle such as dynamical equations, but from the detailed
balance principle, which is
\begin{equation}
\frac{w\left(\vec{x}\rightarrow\vec{x}^{\prime}\right)}{w\left(\vec{x}^{\prime}\rightarrow\vec{x}\right)}
= \frac{e^{-\beta E\left(\vec{x}^{\prime}\right)}}{e^{-\beta
E\left(\vec{x}\right)}}.
\end{equation}
This will give us the transition rate between states $\vec{x}$ and
$\vec{x}^{\prime}$ with the help of other arguments\cite{newman}.
Then, the system can start from an arbitrary initial state, and at
every time step, according to the probability given by the
transition rate, it moves onto or rejects a new test state
randomly chosen from all other states. For a short summery,
whatever the detail of the evolutionary process will be, the
conclusion is, for a system described by Hamiltonian, we do have
some methods to get the thermal equilibrium.

Now we apply this idea onto games. The first requirement is the
system should be described by a Hamiltonian such as a hermitian
operator $\hat{H}$. And the state of system should be represented
by a vector $\vec{x}$ or $\left|\vec{x}\right>$. How far away does
the current formalism of Game Theory leave from those two
requirements? Traditionally, the description of system-level state
does not exist but only a set of states for every single player,
which are a set of probability distributions over their own
strategy spaces. Or, say, in non-cooperative static game, state of
player is $\left\{\rho^{i}\right\}$, where every $\rho^{i}$ is a
probability distribution, such as
\begin{equation}
p^{i}\left(s^{i}_{l}\right) = p^{i}_{l},
\end{equation}
or in our notation,
\begin{equation}
\rho^{i} = \sum_{l}
p^{i}_{l}\left|s^{i}_{l}\right>\left<s^{i}_{l}\right|.
\end{equation}
It's natural to form a system-level distribution for $N$-player
game as
\begin{equation}
P\left(s^{1}_{l_{1}},s^{2}_{l_{2}},\dots, s^{N}_{l_{N}}\right) =
p^{1}_{l_{1}}p^{2}_{l_{2}}\dots p^{N}_{l_{N}}.
\end{equation}
or in our notation,
\begin{equation}
\rho^{S} = \prod_{i=1}^{N}\rho^{i} = \sum_{\left\{l_{i}\right\}}
p^{1}_{l_{1}}p^{2}_{l_{2}}\dots
p^{N}_{l_{N}}\left|s^{1}_{l_{1}},s^{2}_{l_{2}},\dots,
s^{N}_{l_{N}}\right>\left<s^{1}_{l_{1}},s^{2}_{l_{2}},\dots,
s^{N}_{l_{N}}\right|.
\end{equation}
Therefore, the state seems not very far away from the requirement,
at least we already have the vector form as
$\left(s^{1}_{l_{1}},\dots, s^{N}_{l_{N}}\right)^{T}$, or
$\left|s^{1}_{l_{1}},\dots, s^{N}_{l_{N}}\right>$. The main
problem is at the payoff function.

Traditionally, in a $N$-player game, every player is assigned a
payoff function $M^{i}$, which linearly maps $N$ vectors into a
real number, or technically speaking $M^{i}$ is a
$\left(0,N\right)$-tensor. However, an hermitian operator
$\hat{H}$ is required to be a $\left(1,1\right)$-tensor, which
maps a left vector and a right vector into a real number.
Constructing $H^{i}$ from $M^{i}$ will be a significant step in
our formalism. This will be done in section $\S$\ref{manipulate}.
After we get $H^{i}$, we will see it's truly hermitian.

The main task of this paper is to develop a Hamiltonian Formalism
for Game Theory. For classical game, it's mainly from $M^{i}$ to
$H^{i}$. And then we will show the value of such new
representation. Another thread leading to this paper, is to
consider the game-theory problem that when we replace the
classical object with a quantum object, what we need to do for the
general framework? Here the object refers to the object which the
strategies of the game apply on, such as the coin in the Penny
Flipping Game (PFG).

Game Theory\cite{Gamecourse} is to predict the strategy of all
players in a game, no matter a classical object or a quantum
object is used as the game object which strategies acts on. For
example, when we talk about $2$-player PFG, the object is a coin,
which is a classical object with two basic states
$\left\{1(head),-1(tail)\right\}$, so the strategies we can use
are Non-flip and Flip. We denote them as $I, X$ respectively, for
some reason will be known soon. So the strategy state of every
player is a probability distribution over $\left\{I,X\right\}$.
Now, consider a quantum object, a $\frac{1}{2}$-spin instead of a
penny. A spin has more states than a penny. The difference is a
penny only has {\it probability combination} over
$\left\{1,-1\right\}$ states, while a spin has {\it coherent
combination} over $\left\{1,-1\right\}$ states. Or technically, we
say, a general state of a spin is
\begin{equation}
\left|spin\right> = \alpha\left|+1\right> + \beta\left|-1\right>,
\end{equation}
with constrain, $\alpha^{*}\alpha + \beta^{*}\beta = 1$ while a
penny state is a probability distribution such as
\begin{equation}
p\left(+1\right) = p^{+}, p\left(-1\right) = p^{-},
\end{equation}
with constrain, $p^{+} + p^{-} = 1$. It's easier to distinguish
those two objects by using density matrix,
\begin{equation}
\rho^{spin} = \alpha^{*}\alpha\left|+1\right>\left<+1\right| +
\beta^{*}\alpha\left|+1\right>\left<-1\right| +
\alpha^{*}\beta\left|-1\right>\left<+1\right| +
\beta^{*}\beta\left|-1\right>\left<-1\right|,
\end{equation}
while
\begin{equation}
\rho^{penny} = p^{+}\left|+1\right>\left<+1\right| +
p^{-}\left|-1\right>\left<-1\right|.
\end{equation}
The way that we use density matrix to describe state of classical
object recalls the same treatment and understanding mentioned
above as in equ(\ref{pclassical}). We see that for classical
object, at least the density matrix provides equivalent
information with probability distribution. Because of these extra
states, the strategy (operator) space over quantum object is also
larger than the one over the corresponding classical object. Here,
explicitly, for spin, we have four basic strategies (operators):
$I, X, Y, Z$ v.s. $I, X$ for penny. Here $I,X,Y,Z$ refers to
identity matrix, and three Pauli matrixes,
\begin{equation}
\begin{array}{cccc}
I = \left[\begin{array}{cc}1 & 0 \\ 0 & 1\end{array}\right] & X =
\left[\begin{array}{cc}0 & 1 \\ 1 & 0\end{array}\right] & Y =
\left[\begin{array}{cc}0 & -i \\ i & 0\end{array}\right] & Z =
\left[\begin{array}{cc}1 & 0 \\ 0 & -1\end{array}\right]
\end{array},
\end{equation}
It's easy to see that $X$ exchanges the states $\left(+1,
-1\right)$ and $I$ keeps the states. A general quantum operator
can be expressed as
\begin{equation}
s^{i} = e^{i\alpha}\left(\cos{\frac{\gamma}{2}}\cos{\frac{\beta
+\delta}{2}}I + i\sin{\frac{\gamma}{2}}\sin{\frac{\beta
-\delta}{2}}X - i\sin{\frac{\gamma}{2}}\cos{\frac{\beta
-\delta}{2}}Y - i\cos{\frac{\gamma}{2}}\sin{\frac{\beta
+\delta}{2}}Z\right), \label{decompose}
\end{equation} where $\alpha, \beta, \gamma,
\delta$ are real number\cite{qubit}. Therefore, the classical
probability distribution over $I,X$ for classical game should be
replaced by a quantum probability distribution over $I,X,Y,Z$ for
quantum game.

However, there are two ways to do this generalization. One is to
consider a classical probability distribution over all quantum
operators. This was widely used in \cite{meyer, jens, basevec,
entangle}, where the distribution will be a classical probability
combination as,
\begin{equation}
\rho^{player}_{Cl}  =  \sum_{\alpha, \beta, \gamma,
\delta}p^{player}\left(\alpha, \beta, \gamma,
\delta\right)\left|s\right>\left<s\right|.
\end{equation}
We will discuss this in more details in section
$\S$\ref{diffstate}. Another way is a density matrix on the basis
of $\left\{I,X,Y,Z\right\}$, so that
\begin{equation}
\rho^{player}_{Qu} = \sum_{\lambda,
\sigma\in\left\{I,X,Y,Z\right\}}
\rho^{player}_{\lambda\sigma}\left|\lambda\right>\left<\sigma\right|.
\end{equation}
In section $\S$\ref{diffstate}, we will compare those two and
conclude they are different, and hopefully give a conclusion that
the second definition $\rho^{player}_{Qu}$ should be used.

Because the system-level state is our fundamental dynamical
variable, this representation is born as a way to deal with both
non-cooperative and coalitional games. So section
$\S$\ref{coalitional} is devoted to discuss about correlated
strategy state between players. Here correlation refers to both
classically correlation and entanglement in quantum sense. We
mainly want to distinguish the difference between the entanglement
coming from the object states when there are more than one
sub-objects in our quantum game, and the entanglement coming from
correlated players. In that section, we will also show a thread on
the way to generalize our formalism to explicitly include
Coalitional Game Theory.

In the last part of this paper, we come back to our starting
point, whether this new formalism helps to find the Nash
Equilibria. We have not got a general conclusion yet, so just one
representative example is discussed. Also we will put classical
game as a limit situation of our quantum game when the payoff
matrixes $\left\{H^{i}\right\}$ have a special property, namely
``Reducible'' and ``Separable''.

\section{A short historical introduction of Traditional Quantum Game Theory}
\label{history}

Usually, a static and non-cooperative game can be defined as
triple $\left<N,\mathbb{S},G\right>$, in which $N$ is the set of
$N$ players, $\mathbb{S}$ is the strategy space $S^{1}\times
S^{2}\times\cdots S^{N}$ and $S^{i}$ is the $L_{i}$-dimension
strategy space of player $i$, and $G$ is a linear mapping from
$\mathbb{S}$ to N-dimension real space $\mathbb{R}^{N}$. $S^{i}$
is expanded by basic pure strategies $\left\{s^{i}_{l}\right\}$,
where $l=1,2,\dots,L_{i}$. We denote the set of these pure
strategies as $B\left(S^{i}\right)$, which also means the basis of
strategy space $S^{i}$. The difference and relation between
$S^{i}$ and $B\left(S^{i}\right)$ is extremely important in our
future discussion, especially in section $\S$\ref{diffstate}.
Originally, $G$ is only defined on the pure strategies, from
$B\left(S^{1}\right)\times \cdots \times B\left(S^{N}\right)$ to
$\mathbb{R}^{N}$. The idea of mixed strategy game generalizes the
$G$ onto the whole space of $\mathbb{S}$, by probability average
of the pure strategy payoff\cite{Gamecourse}.

Traditionally, in Classical Game Theory, it has never been
explicitly pointed out the relation between strategies and
operators acting on game object, even the existence of such game
object is usually neglected, so the strategies are considered
abstractly and also the space $\mathbb{S}$ is an abstract vector
space. Because only the real numbers in probability distribution
are needed for discussion, such a vector space is considered as a
$N\times L$ dimensional Normalized Euclidian Space
($\mathbb{NR}^{N\times L}$) with constrain $\sum_{l}p^{i}_{l} =
1$, where $p^{i}_{l}$ is the probability of strategy $s^{i}_{l}$.
We know that Quantum Mechanics is about the evolution over
Normalized Hilbert Space ($\mathbb{NH}^{N\times L}$) with
constrain $Tr\left(\rho\right) = 1$. It's easy to show, even just
from our discussion on the relation between classical and quantum
density matrix in the introduction section, $\mathbb{NH}^{N\times
L}$ covers all the structures of $\mathbb{NR}^{N\times L}$, just
like the relation between $\mathbb{C}$ and $\mathbb{R}$.

Because this abstractness of strategies and the similarity between
$\mathbb{NR}^{N\times L}$ and $\mathbb{NH}^{N\times L}$, in
\cite{marinatto}, it has been pointed out that strategy state can
be expressed as vectors in Hilbert space of strategies. The
authors constructed the math form from the experience of Classical
Game. Bra and ket Vector and even Density Matrix has been used to
represent a strategy state there. In \cite{meyer,jens}, it's
emphasized that the strategies should be operators on Hilbert
space of object, not the vector in Hilbert space of object.
Therefore, the starting point of \cite{marinatto} seems wrong
although the theory looks beautiful. According this, the whole set
of quantum operator should be used as strategies, and the strategy
state of a player is still a distribution function over this
strategy space. In \cite{basevec}, it has been noticed that a set
of operators can be chosen as a basis of the operator space. So
that the operator space is also a Hilbert space. But there the
strategy state of a player is still a distribution over the whole
operator space, and now, with base vectors. We name this quantum
game using a probability distribution as the strategy state as
Traditional Quantum Game (TQG). Works in this paper, can be
regarded as a continue along this direction. Our idea is, since
the operator space is also a Hilbert Space, the idea of using
vector and density matrix to represent strategy is correct again,
although at a new level. So in our new formalism, the strategy
state will be a density matrix not a probability distribution.

Because both the classical game and TQG use the probability
distribution although over a larger space, some criticism has been
raised such as in \cite{enk}. In that paper, the authors asked the
problem what's the real difference between so-called quantum game
raised from above papers\cite{meyer,jens,basevec} and the
classical game. Their answer is that no principal difference
exists. Because the new game can be regarded just as a classical
game over a larger space with more strategies. Recalling our penny
and coin example, the only thing changes in the abstract form of
the new game is that a probability distribution over
$\left\{I,X\right\}$ is replaced with another probability
distribution over $\left\{S = \alpha I + \beta X + \mu Y + \nu
Z\right\}$, which has infinite elements. Therefore, the new game
is a infinite-element version of the classical game, there is no
principal difference. We agree with this conclusion, and that's
just why we think our new representation should be used not the
TQG. A density matrix over $\left\{I,X,Y,Z\right\}$ is definitely
different with both classical probability distribution over
$\left\{I,X\right\}$ and the probability distribution over
$\left\{S\right\}$.

In \cite{basevec}, the authors provided another argument about the
value of the TQG. It's about the efficiency. They have claimed
that the efficiency of TQG will be significantly higher than the
corresponding classical one. In this paper, we don't want to
comment on this issue. Because principally, our new representation
is already a new game compared with the classical game. Discussion
about efficiency will be a minor topic later on if necessary.

In \cite{frame,proof}, we have developed the ideas about the
general framework, but mainly from the background of Classical Game
Theory, without a full acknowledgement of the development of TQG.
In those papers, by using the basis of operator(strategy) space, a
full math structure of Hilbert space is applied onto Game Theory,
including Density Matrix for the Strategy State of one player and
all players, Hermitian Operator on the operator(strategy) space as
Payoff Matrix, and also several specific games are analyzed in the
new framework. But we haven't show the difference between our new
quantum game and the TQG proposed in
\cite{meyer,jens,basevec,marinatto}, and neither emphasized on the
privilege of this new representation. In this paper, we will
present our general formalism in a slight different way with more
links with TQG and try to answer those two questions. Also we will
compare our game with the classical game, in the language of our
new representation.

In the next section ($\S$\ref{manipulate}), following the logic
above, first, we want to explicitly point out the relation between
strategies and operators over state space of object. Therefore, a
new concept as Manipulative Definition of game is introduced.

\section{Manipulative definition and abstract definition}
\label{manipulate}

The manipulative definition of static game is a quadruple
\begin{equation}
\Gamma^{m} = \left(\rho^{o}_{0}\in\mathcal{H}^{o}, S \triangleq
\left\{s^{i}\in S^{i}\right\},\mathcal{L},P \triangleq
\left\{P^{i}\right\}\right), \label{mgame}
\end{equation}
and the payoff of player $i$ is determined by
\begin{equation}
E^{i}\left(\rho\right) =
Tr\left(P^{i}\mathcal{L}\left(S\right)\rho^{o}_{0}\mathcal{L}^{\dag}\left(S\right)\right).
\label{mpayoff}
\end{equation}
Let's explain them one by one. $\mathcal{H}^{o}$ is the state
space of object, and $\rho^{o}_{0}$ is its initial state. We
denote this space as $\mathcal{H}^{o}$ because we has shown
previously that even for a classical object, its state in
probability space can be equivalently expressed in Hilbert Space
although no more information is provided. So notation
$\mathcal{H}^{o}$ can cover both classical and quantum objects.
Every $S^{i}$ is $\mathcal{H}^{*}$, the operator space over
$\mathcal{H}^{o}$. $S$ is a combination of $s^{i}\in S^{i}$, and
here $s^{i}$ is a single pure strategy. A generalization from this
definition of pure-strategy game into mixture-strategy game, even
into coalitional game will be done later on by deriving abstract
definition from manipulative definition. $\mathcal{L}$ is a linear
mapping from $S$ to $\mathcal{H}^{*}$ so that it will give an
effective operator over $\mathcal{H}^{o}$ from the combination of
$S$. After we get such an effective operator
$\mathcal{L}\left(S\right)$, the object will transfer to a new
state according to the effect of such operator so that the new
state is
\begin{equation}
\rho^{o}_{new}=
\mathcal{L}\left(S\right)\rho^{o}_{0}\mathcal{L}^{\dag}\left(S\right).
\label{evolve}
\end{equation}
Then $P^{i}$, the linear mapping from $\mathcal{H}^{o}$ to
$\mathbb{R}$ gives a scale to determine the payoff of every
player. All games can be put into this framework. The linear
property of $\mathcal{L}$ and $P^{i}$ is an essential condition of
Game Theory.

The abstract definition mask the existence and information about
the object. It's defined as a dipole
\begin{equation}
\Gamma^{a} = \left(\rho^{S}\in \mathbb{S}\triangleq
\prod_{i=1}^{N} \otimes S^{i},
H\triangleq\left\{H^{i}\right\}\right), \label{absgame}
\end{equation}
and $H^{i}$ is mapping from $\mathbb{S}$ to $\mathbb{R}$ so that
the payoff is determined by,
\begin{equation}
E^{i} = Tr\left(\rho H^{i}\right). \label{abspayoff}
\end{equation}
Compare equ(\ref{mpayoff}) with equ(\ref{abspayoff}), we will see
that $H^{i}$ should include all information implied by
$\rho^{o}_{0}$, $\mathcal{L}$ and $P^{i}$. Besides this, here
$\rho^{S}$, a density matrix state of all players, is also
slightly different with $S$, the strategy combination in the
manipulative definition. While $S$ can only be used in
pure-strategy game, $\rho^{S}$ can be used for general
mixture-strategy game and coalitional game. Before we construct
the relation between $S$ and $\rho^{s}$, and between
$\left(\rho^{o}_{0},\mathcal{L}, P^{i}\right)$ and $H^{i}$, we
want to compare our definition with traditional one first, and
give some examples in above notation.

\subsection{Example of classical game: Penny Flipping Game}
In a $2$-player PFG, each player can choose one from two
strategies $\left\{I,X\right\}$, which represents Non-flip and
Flip operator respectively, and initially, the penny can be in
head state, after the players apply their strategies in a given
order such as player $1$ before player $2$ (here, this order is
not a physical time order but a logic order still happens at the
same time. Sometimes, this order will effect the end state of the
object) the penny transforms onto a new state, and then, the
payoff of each player is determined by the end state of the penny.
For example, player $1$ will win if it's head, otherwise player
$2$ wins. In above notations, the manipulative definition of this
game is,
\begin{equation}
\begin{array}{llll} \rho^{o}_{0} = \left|+1\right>\left<+1\right| =\left[\begin{array}{ll}1 & 0 \\ 0 &
0\end{array}\right], & S^{1} = S^{2}=\left\{I,X\right\}, &
\mathcal{L} = s^{2}s^{1}, & P^{1} = \left[\begin{array}{ll}1 & 0
\\ 0 & -1\end{array}\right]=-P^{2},
\end{array}
\label{pennygame}
\end{equation}
Substituting all above specifics into the general form
equ(\ref{mgame}), it's easy to check it gives the correct payoff
value of each player as described by words above, even when
$s^{1}, s^{2}$ are distributions not single strategies.

The abstract definition is given by
\begin{equation}
H^{1,2} = \left[\begin{array}{llll}1,-1 & 0 & 0 & 0 \\ 0 & -1,1 &
0 & 0 \\ 0 & 0 & -1,1 & 0 \\0 & 0 & 0 & 1,-1
\end{array}\right].
\label{pennyquantum}
\end{equation}
For a comparison, we also give the usual traditional definition
$\Gamma = \left(S,G,N\right)$, where $G$ is the payoff matrix,
\begin{equation}
G^{1,2} = \left[\begin{array}{ll}1,-1 & -1,1 \\ -1,1 & 1,-1
\end{array}\right].
\label{pennyclassical}
\end{equation}
Later on, we will construct explicitly the relation between $P$,
$H$ and $G$. But at current stage, we need to notice that
$\rho_{0},\mathcal{L},P$ in manipulative definition contains
equivalent information with $G$ in traditional abstract definition
and $H$ in our new abstract definition. Furthermore, we will see
the generalization from $G$ to $H$ will give our theory the power
to put both quantum and classical game into the same framework,
and the introduction of the manipulative definition will provide a
very natural way to raise quantum game.

\subsection{Example of quantum game: Spin Rotating Game}
Now, we replace the penny with a quantum spin, which has the
general density matrix states, $\rho^{o} \in \mathcal{H}^{o}$, and
its operators(strategies) form a space $\mathcal{H}^{*}$, whose
basis is $\mathbb{B}\left(\mathcal{H}^{*}\right) =
\left\{I,X,Y,Z\right\}$. Its' manipulative definition is
\begin{equation}
\begin{array}{llll} \rho^{o}_{0} = \left|+1\right>\left<+1\right| =\left[\begin{array}{ll}1 & 0 \\ 0 &
0\end{array}\right], & S^{1} = S^{2}=\mathcal{H}^{*}, &
\mathcal{L} = s^{2}s^{1}, & P^{1} = \left[\begin{array}{ll}1 & 0
\\ 0 & -1\end{array}\right]=-P^{2},
\end{array}
\label{spingame}
\end{equation}
where $\mathcal{H}^{*} = \mathbb{E}\left\{I,X,Y,Z\right\}$ refers
to the Operator Space expanded by base vectors
$\left\{I,X,Y,Z\right\}$. Sometimes, we also denote this relation
as $ \mathbb{B}\left(\mathcal{H}^{*}\right) =
\left\{I,X,Y,Z\right\}$. We call this game as Spin Rotating Game
(SRG). Compared with PFG, the only difference is the set
$\left\{I,X\right\}$ is replaced by
$\mathbb{E}\left\{I,X,Y,Z\right\}$. This replacement is
nontrivial. It's different with a replacement between
$\left\{I,X\right\}$ and $\left\{I,X,Y,Z\right\}$. The later means
the number of pure strategies grows from $2$ to $4$, but in the
former it grows from $2$ to $\infty$ because the operator space
has infinite elements.

The abstract definition is given by
\begin{equation}
H^{1} = \left[\begin{array}{cccccccccccccccc}
1&0&0&1&0&1&-i&0&0&i&1&0&1&0&0&1
\\0&-1&-i&0&-1&0&0&1&-i&0&0&i&0&-1&-i&0
\\0&i&-1&0&i&0&0&-i&-1&0&0&1&0&i&-1&0
\\1&0&0&1&0&1&-i&0&0&i&1&0&1&0&0&1
\\0&-1&-i&0&-1&0&0&1&-i&0&0&i&0&-1&-i&0
\\1&0&0&1&0&1&-i&0&0&i&1&0&1&0&0&1
\\i&0&0&i&0&i&1&0&0&-1&i&0&i&0&0&i
\\0&1&i&0&1&0&0&-1&i&0&0&-i&0&1&i&0
\\0&i&-1&0&i&0&0&-i&-1&0&0&1&0&i&-1&0
\\-i&0&0&-i&0&-i&-1&0&0&1&-i&0&-i&0&0&-i
\\1&0&0&1&0&1&-i&0&0&i&1&0&1&0&0&1
\\0&-i&1&0&-i&0&0&i&1&0&0&-1&0&-i&1&0
\\1&0&0&1&0&1&-i&0&0&i&1&0&1&0&0&1
\\0&-1&-i&0&-1&0&0&1&-i&0&0&i&0&-1&-i&0
\\0&i&-1&0&i&0&0&-i&-1&0&0&1&0&i&-1&0
\\1&0&0&1&0&1&-i&0&0&i&1&0&1&0&0&1
\end{array}\right] = -H^{2}.
\label{spinquantum}
\end{equation}
In the traditional language of Classical Game Theory and also in
the language of TQG (yes, they are the same thing when both refer
to the games on quantum objects), the payoff function for each
player is an infinite-dimension matrix, the corresponding element
when player $1,2$ choose strategy $s^{1,2} = \alpha^{1,2}I +
\beta^{1,2} X + \mu^{1,2}Y + \nu^{1,2}Z$ respectively is,
\begin{equation}
G^{1,2}_{s^{1}s^{2}} =
Tr\left(P^{1,2}s^{2}s^{1}\rho^{o}_{0}\left(s^{1}\right)^{\dag}\left(s^{2}\right)^{\dag}\right),
\label{spinclassical}
\end{equation}
where $P^{1,2}$ and $\rho^{o}_{0}$ are given in the manipulative
definition, equ(\ref{spingame}). Therefore, at last, the payoff
functions will be functions of the parameters
$\alpha^{1,2},\beta^{1,2},\mu^{1,2},\nu^{1,2}$. Later on, we will
know, definitions of classical game equ(\ref{pennygame}),
equ(\ref{pennyquantum}) and equ(\ref{pennyclassical}) are
equivalent, while the definition of quantum game should be in the
form of equ(\ref{spingame}) or equ(\ref{spinquantum}), but not
equ(\ref{spinclassical}).

\subsection{From manipulative definition to abstract definition}
We have given the definitions and the examples, and compared them
with the usual form in the notation of classical games. In a
sense, the manipulative definition is more fundamental because all
its rules are given directly at the object and operator level. Now
we ask the question how to construct the abstract form from the
manipulative form, or in another way, where we got the two
abstract form $H^{1,2}$ for the two games above? And why we think
the two forms give the equivalent description of the same games?

We know the state of object, no matter when it's classical or
quantum object, can be written as density matrix. The idea to
derive abstract form from manipulative form is to represent state
of players by density matrix in the strategy space. In order to do
this, the strategy space has to be a Hilbert Space. We already
know that, in Physics, the object state space $\mathcal{H}^{o}$ is
a Hilbert Space and single-player strategy space is $S^{i} =
\mathcal{H}^{*}$, the operator space over $\mathcal{H}^{o}$. This
operator space with natural inner product as
\begin{equation}
\left<A\right|\left.B\right>\triangleq\left(A,B\right) =
\frac{Tr\left(A^{\dag}B\right)}{Tr\left(I\right)}, \label{inner}
\end{equation}
will become a Hilbert Space. Therefore, the language of density
matrix can be applied in the strategy space now. The total space
of states of all players should be a product space of all such
single-player strategy spaces,
\begin{equation}
\mathbb{S} = \prod_{i=1}^{N}\otimes S^{i}.
\end{equation}
So a state of all players is a density matrix $\rho^{S}$ in
$\mathbb{S}$. Denote the basis of $\mathbb{S}$ as
$\mathbb{B}\left(\mathbb{S}\right)$, then
\begin{equation}
\rho^{S} =
\sum_{\mu,\nu\in\mathbb{B}\left(\mathbb{S}\right)}\rho_{\mu\nu}\left|\mu\right>\left<\nu\right|.
\label{correctstate}
\end{equation}
In a special case of noncooperative game, the density matrix over
the total space $\mathbb{S}$ can be separated as
\begin{equation}
\rho^{S} = \prod_{i=1}^{N}\rho^{i} =
\prod_{i=1}^{N}Tr_{-i}\left(\rho^{S}\right),
\end{equation}
where $Tr_{-i}\left(\cdot\right)$ means to do the trace over all
the other spaces except $S^{i}$ so that $\rho^{i}$ is a density
matrix over $S^{i}$,
\begin{equation}
\rho^{i} = \sum_{\mu^{i},\nu^{i}\in
B\left(S^{i}\right)}\rho^{i}_{\mu^{i}\nu^{i}}\left|\mu^{i}\right>\left<\nu^{i}\right|.
\end{equation}
We have discussed in the section of introduction, state of players
of classical game only need the diagonal terms of this density
matrix because only a probability distribution over strategy space
is needed in classical game, and diagonal terms of density matrix
provide the equivalent structure of probability distribution. So,
why here we need a general density matrix, not a probability
distribution again?

For simplicity, let's discuss this issue in $S^{i}$, a
single-player strategy space, and by the example of SRG, and then
the general counterpart in $\mathbb{S}$ will be very similar. The
basis of $S^{i}$ is $B\left(S^{i}\right) =
\left\{I,X,Y,Z\right\}$. When the player takes strategy $s^{i}\in
S^{i}$, for example, $s^{i} = \frac{1}{\sqrt{2}}\left(X+Y\right)$,
according to our general form, it must be able to express into the
form of density matrix, such as
\begin{equation}
\rho^{i} = \left|s^{i}\right> \left<s^{i}\right| =
\frac{1}{2}\left(\left|X\right> \left<X\right| + \left|Y\right>
\left<Y\right| + \left|X\right> \left<Y\right| + \left|Y\right>
\left<X\right|\right).
\end{equation}
This naturally requires the off-diagonal terms such as
$\left|X\right> \left<Y\right|$. We know that the probability
distribution is used for classical system but density matrix is
used for quantum system. In fact, here we have the same relation.
Probability distribution is good for state of players in games on
classical objects, but density matrix should be used for state of
players in games on quantum objects. In this view point, such
generalization is quite natural. In the next
section($\S$\ref{diffstate}), we will discuss another way of
generalization and compare it with our language of density matrix.

Now, the problem about state has been solved. Next problem should
be how to define Hermitian Operator $H^{i}$ over $\mathbb{S}$ to
give payoff as required in equ(\ref{abspayoff}). The definition
is, for any given basis $\mu,\nu \in
\mathbb{B}\left(\mathbb{S}\right)$,
\begin{equation}
\left<\mu\right|H^{i}\left|\nu\right> =
Tr\left(P^{i}\mathcal{L}\left(\nu\right)\rho^{o}_{0}\mathcal{L}^{\dag}\left(\mu\right)\right).
\end{equation}
Or for a classical game, because the strategy density matrix has
only diagonal term, when we do the trace required in
equ(\ref{abspayoff}), only the diagonal terms of the payoff matrix
will be effective. So for classical game, it can be defined as
\begin{equation}
\left<\mu\right|H^{i}\left|\mu\right> =
Tr\left(P^{i}\mathcal{L}\left(\mu\right)\rho^{o}_{0}\mathcal{L}^{\dag}\left(\mu\right)\right).
\end{equation}
Apply this general definition onto SPG, we get
\begin{equation}
H^{i}_{\mu^{1}\mu^2\nu^{1}\nu^{2}}=\left<\mu^{1}\mu^2\right|H^{i}\left|\nu^{1}\nu^{2}\right>
=
Tr\left(P^{i}\nu^{2}\nu^{1}\rho^{o}_{0}\left(\mu^{1}\right)^{\dag}\left(\mu^{2}\right)^{\dag}\right),
\end{equation}
then $H^{1,2}$ will be a $16\times16$ matrix. And similarly, onto
PFG, we get $4\times4$ matrixes, defined as
\begin{equation}
H^{i}_{\mu^{1}\mu^2\mu^{1}\mu^{2}}=\left<\mu^{1}\mu^2\right|H^{i}\left|\mu^{1}\mu^{2}\right>
=
Tr\left(P^{i}\mu^{2}\mu^{1}\rho^{o}_{0}\left(\mu^{1}\right)^{\dag}\left(\mu^{2}\right)^{\dag}\right).
\end{equation}
It's easy to check $
\left(\left<\mu\right|H^{i}\left|\nu\right>\right)^{*} =
\left<\nu\right|H^{i}\left|\mu\right> $ so that
\begin{equation}
\left(H^{i}\right)^{\dag} = H^{i}.
\end{equation}
Meaning of the diagonal terms such as
$\left<\mu\right|H^{i}\left|\mu\right>$ is just the payoff of
player $i$ when the state of all players is $\left|\mu\right>$,
however, the meaning of the off-diagonal terms is not such
straightforward. For games on quantum objects, generally the
off-diagonal terms are necessary. For example, in SRG, when the
basis $\left\{I,X,Y,Z\right\}$ is used, $s = \frac{X+Y}{\sqrt{2}}$
is also a pure strategy, so we will need to consider the payoff
when both players choose this pure strategy. This is
\begin{equation}
\begin{array}{ccc}
\left<ss\right|H^{i}\left|ss\right> & = &
\frac{1}{4}\left<\left(X+Y\right)\left(X+Y\right)\right|H^{i}\left|\left(X+Y\right)\left(X+Y\right)\right>
\\
\\ & = &
\frac{1}{4}\left(\left<XX\right|H^{i}\left|XX\right> +
\left<XX\right|H^{i}\left|XY\right> + \cdots
\left<YY\right|H^{i}\left|YY\right>\right)
\end{array},
\end{equation}
totally $16$ terms including the off-diagonal terms such as
$\left<XX\right|H^{i}\left|XY\right>$. For games on classical
objects, it's always possible to find a basis, under which all
$H^{i}$s have only the diagonal terms. But for games on quantum
objects, generally, it's not. In section $\S$\ref{reduction}, we
will discuss this in more details, try to find the condition when
quantum games can be reduced into classical games.

Now, our general definition of both quantum and classical games
has been given, starting from the manipulative definition. In
fact, in some sense, this is a generalization from manipulative to
abstract definition, not just a transformation. In the former,
only non-cooperative players are allowed, because the whole state
$\rho^{S}$ is a product of single-player state $\rho^{i}$, but in
the later, a general non-direct-product density matrix
$\rho^{S}\neq\prod_{i=1}^{N}\rho^{i}$ can be used as state of
players. This will include the coalitional games (where the
players are correlated by the way of classical correlated
probability distribution) into the same framework, even also
including quantum correlated players. This is different with the
correlation coming from the classical or quantum objects. The
details will be discussed in $\S$\ref{coalitional}.

In order to give a complete definition of Game Theory, we need to
redefine the Nash Equilibrium (NE) for both classical and quantum
games, and generalize it a little onto games with state of players
as a non-direct-product density matrix. NE is redefined as
$\rho^{S}_{eq}$ that
\begin{equation}
E^{i}\left(\rho^{S}_{eq}\right) \geq
E^{i}\left(Tr^{i}\left(\rho^{S}_{eq}\right)\rho^{i}\right),
\forall i, \forall \rho^{i}. \label{newne}
\end{equation}
The special case of this definition, when we consider
non-cooperative classical game, naturally becomes the usual NE,
$\rho^{S}_{eq} =
\left(\rho^{1}_{eq}\cdots\rho^{i}_{eq}\cdots\rho^{N}_{eq}\right)$,
that
\begin{equation}
E^{i}\left(\rho^{1}_{eq}\cdots\rho^{i}_{eq}\cdots\rho^{N}_{eq}\right)
\geq
E^{i}\left(\rho^{1}_{eq}\cdots\rho^{i}\cdots\rho^{N}_{eq}\right),
\forall i, \forall \rho^{i}. \label{oldne}
\end{equation}
This should be easily recognizable for readers from Game Theory.
Now our goal of Game Theory becomes to find the NE $\rho^{S}_{eq}$
as in equ(\ref{newne}) in the strategy space of $\mathbb{S} =
\prod_{i=1}^{N}\otimes S^{i}$ given payoff matrixes $H =
\left\{H^{i}\right\}$ as in equ(\ref{absgame}).

It seems at first sight, even for classical game, this
representation is much more complex than the traditional one. But,
we think this step is necessary for the development of Game
Theory, first, because this new representation is a system of
Hamiltonian Formalism so that it's easy to develop the Statistical
Physics upon this; second, games on quantum objects require such
formalism.

\section{Compared with TQG}

From section $\S$\ref{history}, we know there is another kind of
Quantum Game Theory, TQG. Is it an equivalent theory just another
representation, or different with ours? We want to compare those
two in two aspects: first, the definition of state of all players;
second, the effect of the correlation coming from quantum objects.

\subsection{Strategy state: density matrix vs. probability distribution}
\label{diffstate}

Let's start from the strategy state of a single quantum player.
There are two different possible forms. Recalling the manipulative
definition of the two examples, PFG in equ(\ref{pennygame}) and
SRG in equ(\ref{spingame}), the only difference is a
finite-element classical pure strategy set $\left\{I,X\right\}$ is
replaced with an infinite-element quantum pure strategy set
$\mathbb{E}\left\{I,X,Y,Z\right\}$. Therefore, a natural
generalization is to set the state of quantum player as a
distribution over the quantum operator space $S^{i} =
\mathbb{E}\left\{I,X,Y,Z\right\}$ as
\begin{equation}
\rho^{i,q} = \sum_{s^{i} \in
S^{i}}p^{i,q}\left(s^{i}\right)\left|s^{i}\right>\left<s^{i}\right|.
\label{wrongsinlestate}
\end{equation}
So a strategy state of all players can be written as
\begin{equation}
\rho^{S,q} = \sum_{S \in
\mathbb{S}}p^{S,q}\left(S\right)\left|S\right>\left<S\right|,
\label{wrongstate}
\end{equation}
where $S = \left|s^{1},s^{2},\cdots, s^{N}\right>$. The meaning of
this kind of state is a probability distribution over a larger
space is used to replace the counterpart in the classical game.
But principally, there is nothing new in the sense of Game Theory.
Such trivial generalization also neglects an important
information, that $s^{i}$ is element in space expanded by
$\left\{I,X,Y,Z\right\}$, so that there is a relation between
$s^{i}$ and the basis $\left\{I,X,Y,Z\right\}$ as in
equ(\ref{decompose}). If we make use of the decomposition in
equ(\ref{decompose}), then we can denote the same relation by a
density matrix form, with $16$ terms,
\begin{equation}
\left|s^{i}\right>\left<s^{i}\right| =
\cos^{2}{\frac{\gamma}{2}}\cos^{2}{\frac{\beta
-\delta}{2}}\left|I\right>\left<I\right| + \cdots +
\cos^{2}{\frac{\gamma}{2}}\sin^{2}{\frac{\beta
-\delta}{2}}\left|Z\right>\left<Z\right|.
\end{equation}
And generally, any single-player states of strategy can be
expressed similarly by
\begin{equation}
\rho^{i} = \sum_{\mu,\nu\in
B\left(S^{i}\right)}\rho^{i}_{\mu\nu}\left|\mu\right>\left<\nu\right|.
\label{correctsinglestate}
\end{equation}
Therefore, a general strategy state of all players should have the
form as in equ(\ref{correctstate}), a density matrix over the
whole strategy space $\mathbb{S}$ expanded by its basis
$\mathbb{B}\left(\mathbb{S}\right)$.

Now, the problem is whether the definition of
equ(\ref{correctsinglestate}) is equivalent with
equ(\ref{wrongsinlestate}), or not? The answer is definitely not.
The normalization conditions from the former and the later are
\begin{equation}
\left\{ \begin{array}{c} \sum_{\mu \in
B\left(S^{i}\right)}\left<\mu\right|\rho^{i}\left|\mu\right> = 1
\\
\\
\sum_{s^{i}\in
S^{i}}\left<s^{i}\right|\rho^{i,q}\left|s^{i}\right> = 1
\end{array}\right.,
\end{equation}
respectively. The difference can be seen from the following
example.
\begin{equation}
\rho^{i} = \frac{1}{4}\left(\left|X+Y\right>\left<X+Y\right| +
\left|X-Y\right>\left<X-Y\right|\right)
\end{equation}
is a good single-player state when the first normalization is
applied, but it's not a good state under the second normalization
condition. This is easy to see if we apply the summation among
$X,Y,\frac{X+Y}{\sqrt{2}}, \frac{X-Y}{\sqrt{2}}$, the result is
$\frac{5}{4}>1$. Readers from TQG may argue that even in the sense
of strategy state of the probability distribution, the density
matrix above should be recognized only as a probability
distribution over $\frac{X+Y}{\sqrt{2}}, \frac{X-Y}{\sqrt{2}}$ so
that the total probability is still $1$. In fact, this implies a
very strong assumption that
$\left<s_{1}^{i}\left|\right.s_{2}^{i}\right> = 0\left(\forall
s_{1}^{i}\neq s_{2}^{i} \right)$, whatever the real forms of them
are. First, this assumption is totally incompatible with
equ(\ref{inner}), our definition of inner product over operator
space. Second, even under such assumption, looking at the same
example, we will find that this is good in the sense of
probability distribution but not good in the sense of density
matrix (normalized to $0$ when $B\left(S\right) =
\left\{I,X,Y,Z\right\}$ is used). So the endogenous relation among
all strategies (operators) and the assumption of orthogonal
relation among them are incompatible.

Now, since they are different, which definition should be used? We
prefer the density matrix over
$\mathbb{B}\left(\mathbb{S}\right)$, not the probability
distribution over $\mathbb{S}$. The first reason is given above
that the former one keep the endogenous relation between
operators. The second reason is even from the practical point
view, discussion in density matrix form is much easier in
probability distribution form as we will see in section
$\S$\ref{reduction} when both of them are applied onto SRG.

\subsection{Entangled strategy state: correlation between players}
\label{coalitional}

Besides the definition of the new quantum game, another privilege
of the new representation is it can be used for coalitional game.
An investigation of the detail correspondence between the new
representation and the coalitional game is still in progress, but
compared with the classical payoff $(0,N)$-tensor, which can only
be used for non-cooperative game, the new representation can be
still be used when the players are not independent. In the
traditional language, the state of all players is determined by a
combination of all single-player state, which is a distribution
over its own strategy space. But in our representation, a density
matrix over the whole strategy space $\mathbb{S}$ is used. So
principally it can be non-direct-product state, so that
\begin{equation}
\rho^{S} \neq \prod_{i=1}^{N}Tr_{-i}\left(\rho^{S}\right),
\label{dependent}
\end{equation}
In Quantum Mechanics, such state is named as an Entangled State.
In our case, because, this entanglement is not in the state space
of the object, but the strategy space of game players, we call
this Entangled Strategy State. This entanglement has two styles.
First, for a classical game, in fact, it means a classical
correlation between players. Second, it can also be the
entanglement between quantum players as in quantum game.

However, no matter which game it refers to, the entanglement is
NOT the usual entangle states of the object, while currently the
most commonly accepted meaning of Entangled Game in TQG does come
from the entanglement of the initial states of the quantum
objects\cite{entangle,enk}. In TQG, it refers to the situation
when more than one sub-objects are used as the object of the
quantum game. For example, two spins are used as the object. Then,
the initial state, $\rho^{o}_{0}$, in our manipulative definition
can be an entangled state. This effect coming from the
entanglement at object level is so-called Entangled Quantum Game.
We have to point out that this entanglement will not lead to an
entanglement between players. Only the correlation between players
will significantly effects the structure of the game, not the
entanglement at object level. When the $\rho^{o}_{0}$ is changed,
$H^{i}$ will changes together, but nothing need to do about
$\rho^{S}$. By our representation, we have proposed an artificial
quantum game, where the global optimism solution is an entangled
state between players\cite{realentangle}. In that paper, we
proposed that when $\rho^{S}$ is not a direct product state, but
\begin{equation}
\rho^{S} = \prod_{\alpha_{i}}\rho^{\alpha_{i}},
\end{equation}
where $\left\{\alpha_{i}\right\}$ is not a set of single players
but a set of subsets totally covering the whole set of players.
Then, any subset $\alpha_{i}$ can be regarded as a coalition of
players. Therefore, a structure of core\cite{Gamecourse} in
Coalitional Game can be very naturally raised from our general
framework of Game Theory. Generally, if we can get the general NEs
defined in equ(\ref{newne}), decompose it as far as possible, the
subset $\alpha_{i}$ will naturally forms a structure of core. The
investigation on the equivalence between this framework and
Coalitional Game is still in progress.

\subsection{Redundant operators from density matrix form}
\label{redundant}

There is one serious problem of this density matrix representation
of operator: although every unitary operator can be expressed as a
density matrix form, not every density matrix corresponds to a
unitary operators or a probability combination of unitary
operators. In Quantum Mechanics, this is not a problem. Every pure
state can be described by a density matrix, and every density
matrix is a pure state or a probability combination of pure
states. The difference between here Quantum Game and Quantum
Mechanics is: the general pure strategy is the expansion of
equ(\ref{decompose}), where only three real variables are used,
while a general pure state is a superposition of base vectors by
complex variables. For example, $\left(X+iY\right)$ will not give
us a unitary operator thus not a good strategy, but
$\left(\left|\uparrow\right>_{z} +
i\left|\downarrow\right>_{z}\right)$ is a good pure state.

Usually, in quantum world, only unitary operators can be used to
manipulate quantum objects. Because of this, it's possible that
when we find $\rho_{eq}^{S}$, but it in fact is a nonapplicable
NE. Two ways may take us out of this problem. One, non-unitary
operator can also be used to manipulate quantum objects. For
example, if quantum measurement is one of the strategies we can
choose, then non-unitary operators are also acceptable as
strategies. Second, restrict our density matrix to allow only
unitary operators. Specifically, for $SU\left(2\right)$, we have
such way of restricted density matrix. From equ(\ref{decompose}),
if a new basis as $\left\{I,iX,iY,iZ\right\}$ is used, the
coefficients in the decomposition will be real numbers. Under this
new basis, if we restrict elements as real numbers,
$\rho^{i}_{\mu\nu}\in \mathbb{R}$, then every density matrix
corresponds to a combination of unitary operators. A general such
restricted density matrix has $3+6=9$ independent real elements. A
probability combination of four orthogonal unitary operators has
$3$ real number for distribution probability, $3$ real number for
the first arbitrary unitary operators, $2$ real number for the
second one, and $1$ real number for the third one, and $0$ for the
last one, so totally also $3+3+2+1=9$ real variables. Generally,
for $SU\left(N\right)$, similar condition of restriction can be
got: finding the decomposition relation similar with
equ(\ref{decompose}), choosing special basis so that all the
coefficients are real, and then restricting the density matrix
only has real elements.

\section{Application of this hamiltonian formalism onto specific games}
We have shown that for a classical game, the language of density
matrix and hermitian operators gives at least the equivalent
description and the TQG should be replaced by our new quantum
game. Now, we consider the following problems. First, if there is
any privilege that a density matrix language will help to solve
the problems in Classical Game? Second, whether there are some
examples of games, showing that the new quantum game should be
used instead of the TQG. In the following two sub sections, we
will answer them one by one.

\subsection{Evolutionary process leading to NEs}
\label{evolution}

The first problem is actually our starting point of this paper,
although in fact, we get some results in general representation
far away from this point. The motivation driving us alone the way
of density matrix and hermitian operators is the hope that this
new formalism will help to construct an evolutionary process
leading to NEs from arbitrary initial strategy state.

During our revision of this paper, we noticed that the idea of
applying Boltzman Distribution and Metropolis Method onto Game
Theory has been proposed in \cite{boltzman}. There the author
suggests a very similar procedure and gives the parameter $\beta$
an explanation as a measurement of the rationality of players.
Even the equ(3) in that paper is exactly the same meaning with our
equ(\ref{boltzmanevolution}). However, the algorithm problem,
although the starting point, is not our central problem. We start
our work from it but end up at a general representation. So it's a
hen for our work, and hopefully it lays a golden egg here, doesn't
it? The only difference is that \cite{boltzman} uses the language
of classical game, so probability distribution and payoff function
are used there, while here, we use the language of density matrix
and hermitian operator. The hamiltonian formalism and operator
form is easier to be generalized onto quantum system, where the
quantum state has to be a density operator. Because of this, here
we still introduce the operator form of this evolutionary
procedure.

The idea is to make use of the Boltzman Distribution as
equ(\ref{boltzman}) to represent the density matrix state over the
strategy space. Because in our language the payoff matrixes now
are hermitian operators, so $e^{\beta H^{i}}$ does look like a
density matrix. The only problem is that usually $H^{i} \neq
H^{j}$, so there is not a system level payoff matrix, while for a
physical system, such a system level hamiltonian always exists.
Therefore, even we generalize the NE onto a general game not
necessary non-cooperative, but we don't have the way to get such
general solution, unless for a special game, where $H^{i}=H^{j}$.
But such game is trivial. It's not game any more, but an
optimization problem. So at current stage, our new representation
help us nothing about a general game. Of course, we believe this
is not because of the representation, but because we haven't found
the right algorithm yet. So coming back to the starting point, how
would it help on the non-cooperative game, where $\rho^{S} =
\prod_{i}\rho^{i}$? Every single-player state $\rho^{i}$ can
evolve in its own space, and is related each other just by
$H^{i}$, which depends on all-player state not only $\rho^{i}$.
The following evolution realizes this picture.
\begin{equation}
\rho^{i}\left(t+1\right) = e^{\beta H^{i}\left(t\right)_{R}},
\label{boltzmanevolution}
\end{equation}
where $H_{R}^{i}$ is the reduced payoff matrix, which means the
payoff matrix when all other players' states are fixed. It's
defined as
\begin{equation}
H^{i}_{R} =
Tr_{-i}\left(\rho^{1}\cdots\rho^{i-1}\rho^{i+1}\cdots\rho^{N}H^{i}\right).
\end{equation}
Payoff value can also be calculated by the reduced payoff matrix
as
\begin{equation}
E^{i} = Tr^{i}\left(\rho^{i}H^{i}_{R}\right) \equiv
Tr\left(\rho^{1}\cdots\rho^{i-1}\rho^{i}\rho^{i+1}\cdots\rho^{N}H^{i}\right).
\end{equation}
Now we have the way to get NEs from arbitrary initial strategy:
\begin{equation}
\rho^{1}\left(0\right)\cdots\rho^{N}\left(0\right) \rightarrow
H^{1}_{R}(0) \rightarrow \rho^{1}\left(1\right) \rightarrow
H^{2}_{R}(1) \rightarrow \rho^{2}\left(1\right) \rightarrow \dots
\rightarrow \rho^{1}_{fix}\cdots\rho^{N}_{fix}.
\end{equation}
The general proof of the existence of the fixed points and the
equivalence between such fixed points with NEs is in progress. In
this paper, we just show the applicable value of this evolutionary
process by one example, Prisoner's Dilemma, where
\[H^{1,2} =
\left[\begin{array}{cccc}-2,-2 & 0 & 0 & 0 \\0 & -5,0 & 0 & 0 \\ 0
& 0& 0,-5 & 0\\0 & 0 & 0 & -4,-4 \end{array}\right].
\]
More examples and discussion about the
stability of the fixed points and its relation with refinement of
NEs can be found in \cite{frame}.

\begin{figure}
\includegraphics{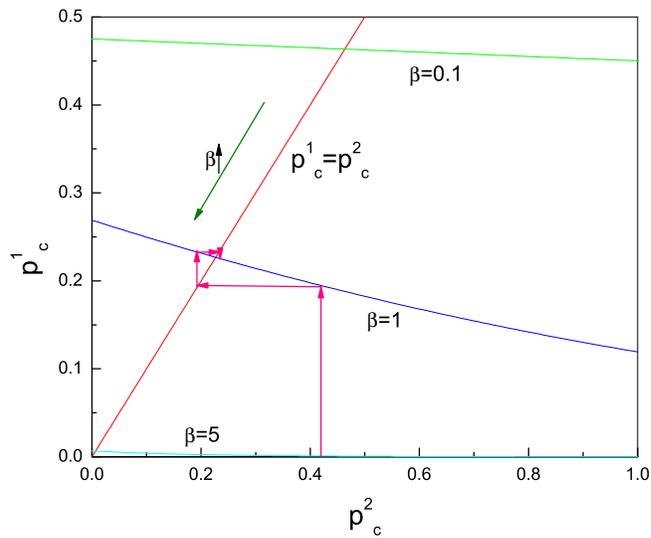}
\caption{The evolutionary procedure applied onto a classical game
with single NE, Prisoner's Dilemma. The
iteration process defined by equ(\ref{boltzmanevolution}) with
different value of parameter $\beta$, ends up at one fixed point,
which decreases from a finite number to 0 when $\beta$ grows. The
meaning of parameter $\beta$ is a measurement of rationality of
players\cite{boltzman}, and $\beta = \infty$ corresponds to full
rational, the case assumed in NE. The solution $0,0$ means both
player choose the second strategy and get $-4,-4$ respectively at
the end. This gives the same NE as in Classical Game Theory.}
\label{graphprisoner}
\end{figure}

\subsection{NEs of SRG}
In this section, we will discuss the non-cooperative Nash
Equilibria of the Spin Rotating Game, defined in
equ(\ref{spingame}) with payoff Hamiltonian
equ(\ref{spinquantum}). According to the restriction discussed in
$\S$\ref{redundant}, basis $\left\{I,iX,iY,iZ\right\}$ is better
to be used here. The payoff Hamiltonian will transform
consistently. Density matrix of single player state is describe
$9$ real numbers, so the strategy state of the two players is
\begin{equation}
\rho^{q,S} = \left[\begin{array}{llll}p^{1}_{11} & \alpha^{1} &
\beta^{1} & \gamma^{1}
\\
\alpha^{1} & p^{1}_{22} & \mu^{1} & \nu^{1}
\\
\beta^{1} & \mu^{1} & p^{1}_{33} & \delta^{1}
\\
\gamma^{1} & \nu^{1} & \delta^{1} & p^{1}_{44}
\end{array}\right]\otimes\left[\begin{array}{llll}p^{2}_{11} & \alpha^{2} &
\beta^{2} & \gamma^{2}
\\
\alpha^{2} & p^{2}_{22} & \mu^{2} & \nu^{2}
\\
\beta^{2} & \mu^{2} & p^{2}_{33} & \delta^{2}
\\
\gamma^{2} & \nu^{2} & \delta^{2} & p^{2}_{44}
\end{array}\right],
\label{sfgstate}
\end{equation}
with the constrain
$p^{1,2}_{11}+p^{1,2}_{22}+p^{1,2}_{33}+p^{1,2}_{44}=1$. Then the
payoff is a function of this $18$ real variables,
\begin{equation}
\begin{array}{lll}
E^{1} = & \begin{array}{l}1 - 2 p^1_{22} - 2 p^1_{33} - 2 p^2_{22}
- 2 p^2_{33} \\+ 4 p^1_{22} p^2_{22} + 4 p^1_{22} p^2_{33} + 4
p^1_{33} p^2_{22} + 4 p^1_{33} p^2_{33}\\ - 4 \alpha^1 \alpha^2 -
4 \beta^1 \beta^2 + 4 \nu^1 \nu^2 + 4 \delta^1 \delta^2 \\ + 4
\alpha^1 \delta^2 - 4 \delta^1 \alpha^2 + 4 \nu^1 \beta^2 - 4
\beta^1 \nu^2
\end{array}
& =-E^{2}.
\end{array}
\end{equation}
Next step, we need to solve all NEs from these two functions. One
way to find some, but not all, of them is to solve the equations,
$\frac{\partial}{\partial x^{2}}E^{1} = 0
=\frac{\partial}{\partial x^{1}}E^{2}$. Here we still don't have
the general algorithm for quantum game, so in order to demonstrate
how our general representation can be used to discuss quantum game
in some practical senses, we just use this trivial way to get some
NEs. The equations gives condition on the possible value of all
$18$ variables as following,
\begin{equation}
\left\{
\begin{array}{l}
p^{1}_{22}+p^{1}_{33}=\frac{1}{2}=p^{2}_{22}+p^{2}_{33}
\\
\alpha^{1}+\delta^{1}=0=\alpha^2-\delta^2
\\
\beta^{1}-\nu^{1} = 0 = \beta^{2} + \nu^{2}
\end{array}
\right.
\end{equation}
Therefore, the NE strategy states could be
\begin{equation}
\rho^{q,S} = \left[\begin{array}{llll}p^{1}_{a} & \alpha^{1} &
\beta^{1} & \gamma^{1}
\\
\alpha^{1} & p^{1}_{b} & \mu^{1} & \beta^{1}
\\
\beta^{1} & \mu^{1} & \frac{1}{2}-p^{1}_{b} & -\alpha^{1}
\\
\gamma^{1} & \beta^{1} & -\alpha^{1} & \frac{1}{2}-p^{1}_{a}
\end{array}\right]\otimes\left[\begin{array}{llll}p^{2}_{a} & \alpha^{2} &
\beta^{2} & \gamma^{2}
\\
\alpha^{2} & p^{2}_{b} & \mu^{2} & -\beta^{2}
\\
\beta^{2} & \mu^{2} & \frac{1}{2}-p^{2}_{b} & \alpha^{2}
\\
\gamma^{2} & -\beta^{2} & \alpha^{2} & \frac{1}{2}-p^{2}_{a}
\end{array}\right].
\label{sfgne}
\end{equation}
For example, the classical sub-game with an NE $p^{1}_{a} =
\frac{1}{2}=p^{2}_{a}$, is covered by this general NE of quantum
game. The intuitive meaning of this general NE can be revealed
partially by a special case such as $p^{1}_{a} =
\frac{1}{4}=p^{1}_{b}, \alpha^{1}\in\left[0,\frac{1}{4}\right]$,
which gives the strategies as following,
\begin{equation}
\left\{
\begin{array}{ll}
s^{1}_{1} =\frac{-I+iX}{\sqrt{2}}; & p^{1}_{1} =
-\alpha^1+\frac{1}{4}\\
s^{1}_{2} =\frac{iY+iZ}{\sqrt{2}}; & p^{1}_{2} =
-\alpha^1+\frac{1}{4}\\
s^{1}_{3} =\frac{I+iX}{\sqrt{2}}; & p^{1}_{3} =
\alpha^1+\frac{1}{4}\\
s^{1}_{4} =\frac{iY-iZ}{\sqrt{2}}; & p^{1}_{4} =
\alpha^1+\frac{1}{4}
\end{array}\right..
\end{equation}
This means a combination of this four pure strategies can act as
the mixture strategy of player $1$ in the equilibria.

Works towards an algorithm for the complete NEs and for general
games is still in progress. However, by this example, we have
shown that our general formalism for sure can be used to describe
the general quantum game. And only finite real variables are
required, not like the situation in TQG, where infinite number of
parameters are needed for general mixture strategy NEs.

\subsection{From Quantum Game to Classical Game}
\label{reduction}

A sub-game of this quantum SRG can be defined as the game
restricted only in pure and mixture combination of strategies
$I,iX$. One may guess that this sub-game will be like the
classical PFG, so the payoff matrix should like
equ(\ref{pennyquantum}). However, the corresponding sub-matrix, in
fact, is
\begin{equation}
H_{sub}^{1,2} = \left[\begin{array}{llll}1,-1 & 0 & 0 & 1,-1 \\ 0
& -1,1 & -1,1 & 0 \\ 0 & -1,1 & -1,1 & 0 \\1,-1 & 0 & 0 & 1,-1
\end{array}\right].
\label{subgame}
\end{equation}
Only the diagonal terms agree with the payoff matrix in
equ(\ref{pennyquantum}), but it also has more non-zero
off-diagonal terms. This raises a question on the condition when a
quantum game, or a general game given in the manipulative
definition as in equ(\ref{manipulate}), becomes a classical game.

The first case is when only the diagonal part of payoff matrix is
chosen as the payoff matrix of a classical game. This situation
happens when we restrict the strategy density matrix only as
probability combination of some fixed and given pure strategies.
In fact, the relation between above sub-game and PFG is such a
case. Under such situation, the set of pure strategies has to be a
priori given.

The second case is when a special basis can be found, that under
this basis both all $\rho^{i}$ and all $H^{i}$ are diagonal. This
requires two facts, first, all payoff matrix commutate with each
other,
\begin{equation}
\left[H^{i},H^{j}\right] = 0, \forall i,j;
\end{equation}
second, the common eigenstates are in the direct-product form, not
entangled,
\begin{equation}
\left|E_{m}\right> =\left|E^{1}_{m}\right>\left|E^{2}_{m}\right>.
\end{equation}

We have an example of the first case, as SRG and PFG. However,
usually, we don't have an concrete example where the requirement
in the second case is fulfilled. For SRG, although
$\left[H^{1},H^{2}\right] = 0$, the common eigenstates are
entangled. The two requirement in the second case will be
fulfilled if the payoff matrices are in the direct-product form:
$H^{i} = h^{i,1}\otimes h^{i,2}$ and $\left[h^{i,m},h^{j,m}\right]
= 0$. But we don't know if there is a real game existing in such a
form. Therefore, we'd better adopt the first case as the way from
quantum game to classical game: only the probability combination
of given pure strategies can act as strategy in the game so that
only the diagonal part of payoff matrix is effective.

\section{Remarks}
Both Classical Game and Quantum Game, both non-cooperative game
and coalitional game, can be generally unified in the new
representation, the operator representation, or the Hamiltonian
Formalism. We have shown that the new quantum game defined in this
new representation is different with the Traditional Quantum Game,
and is not in the scope of Classical Game Theory, which usually
uses the probability distribution over the set of pure strategies
as strategy state. And the entanglement of strategy is not the
usual meaning of entanglement of object state.

The classical game and the TQG are just the same thing, except in
the latter, a lager strategy space has to be used. That's the
reason of the conclusion in \cite{enk} and \cite{basevec}, it was
claimed that, there is no independent meaning of quantum game,
only possibly with different efficiency. However, our quantum game
is different with TQG. The probability distribution is replaced by
a density matrix. And this replacement is significantly different.
So, if we compare the classical game with our new quantum game,
the difference is: both $\rho^{C}$ and $H^{i,C}$, state and payoff
matrixes of classical players are diagonal matrix, while the ones
of quantum player $\rho^{Q}$ and $H^{i,Q}$ generally have
off-diagonal elements. This is similar to the difference between
classical system and quantum system. Therefore, our new quantum
game has independent meaning other than the classical game, just
like the relation between Classical Mechanics and Quantum
Mechanics, but under the same spirit of Game Theory. By replacing
the TQG with our new quantum game, the spirit of Game Theory is
reserved while the scope of game theory is extended. So we are not
arguing that the TQG is independent of classical game, but our new
quantum game is. The operators $H$ used in abstract definition
looks similar with the matrix form used in \cite{basevec},
however, with different meanings. In our notation, $H$ are
operators which give payoff when acting on state of density
matrix, while in \cite{basevec} they does not have the meaning of
an operator over a Hilbert space, and the state of players there
is still a probability distribution.

It's argued in section $\S$\ref{reduction}, principally and by one
example, lower work load need to be done to search for the NEs in
our representation than in the traditional quantum game. However,
neither the existence of the general/special NEs been proved here,
nor an applicable algorithm to find them has been discovered. The
relation between our general NEs and Coalitional game has been
reached by examples in \cite{realentangle}, but not treated
generally and theoretically.

Another thing need to be mentioned is the necessity to introduce
the manipulative definition. At one hand, in fact, it's not
necessary. Here, it's introduced to be a bridge between classical
and quantum game. From the manipulative definition, it's easy to
know the quantum games are just games on quantum objects while
classical games are games on classical objects. At the other hand,
it's valuable. Readers from Classical Game Theory usually work on
games directly at the abstract definition. But we believe that
every classical game can be mapped onto a manipulative definition,
although probably not unique. For example, one can always check
the background where the game comes from. Whenever we have the
manipulative definition, quantization of such game will be easy.

\section{Acknowledgement}
Thanks Dr. Shouyong Pei for the stimulating discussion during
every step of progress of this work. Thanks Dr. Qiang Yuan, Dr.
Zengru Di and Dr. Yougui Wang for their help on the entrance of
Game Theory, their great patience to discuss with me when the idea
of this paper is so ambiguous and rough and also for their
invitation so that I can get the chance to give a talk and discuss
with more people on this. Thanks should also be given to Prof.
Zhanru Yang and Dr. Bo Chen for their reading and comments, and
Dr. Jens Eisert for his encouragement. Thanks Taksu Cheon for the
discussion on the mapping between strategy vector and density
matrix.

\end{document}